\documentclass[runningheads]{llncs}
\usepackage[T1]{fontenc}
\usepackage{color}

\usepackage[table,xcdraw]{xcolor}

\usepackage{graphicx}
\usepackage[utf8]{inputenc}
\usepackage[table,xcdraw]{xcolor}
\usepackage{algorithm}
\usepackage{algpseudocode}
\usepackage{amsmath}
\usepackage{float}
\usepackage{tcolorbox}
\usepackage[utf8]{inputenc}
\usepackage[table,xcdraw]{xcolor}
\usepackage{tikz}
\usepackage[hyphens]{url}
\usepackage{hyperref}
\usepackage{tcolorbox}
\usetikzlibrary{shapes.geometric, arrows}

\newenvironment{takeawaybox}[1]
  {\begin{tcolorbox}[colframe=gray, colback=white, title=#1]}
  {\end{tcolorbox}}

\newcommand{\graycircle}[1]{\textbullet \hspace{#1pt}}

\tikzstyle{box} = [rectangle, rounded corners, minimum width=5.5cm, minimum height=1.5cm, text centered, draw=black, fill=white, thick]
\tikzstyle{arrow} = [thick, ->, >=stealth]

\usepackage{listings}

\lstdefinestyle{algorithmstyle}{
    backgroundcolor=\color{gray!10}, 
    basicstyle=\ttfamily\footnotesize, 
    frame=single, 
    keywordstyle=\bfseries\color{blue}, 
    commentstyle=\itshape\color{green!50!black}, 
    stringstyle=\color{red}, 
    numbers=left, 
    numberstyle=\tiny\color{gray}, 
    stepnumber=1, 
    numbersep=5pt, 
    showstringspaces=false, 
    tabsize=2, 
    breaklines=true, 
    breakatwhitespace=false, 
    captionpos=b 
}
\begin{document}
\title{Large Language Models for Code Generation: The Practitioners’ Perspective}

\author{
    Zeeshan Rasheed\inst{1} \and  
    Muhammad Waseem\inst{1} \and
    Kai Kristian Kemell\inst{1} \and
    Aakash Ahmad\inst{2} \and
    Malik Abdul Sami\inst{1} \and
    Jussi Rasku\inst{1} \and
    Kari Systä\inst{1} \and
    Pekka Abrahamsson\inst{1}
}

\authorrunning{Z. Rasheed et al.}

\institute{
    Faculty of Information Technology and Communication Sciences, Tampere University, 33014 Tampere, Finland \\
    \email{zeeshan.rasheed@tuni.fi, muhammad.waseem@tuni.fi, kai-kristian.kemell@tuni.fi, malik.sami@tuni.fi, jussi.rasku@tuni.fi, kari.systa@tuni.fi, pekka.abrahamsson@tuni.fi} \and
    School of Computing and Communications, Lancaster University Leipzig, Germany \\
    \email{a.ahmad13@lancaster.ac.uk}
}

\maketitle              
\begin{abstract}

Large Language Models (LLMs) have emerged as coding assistants, capable of generating source code from natural language prompts. With the increasing adoption of LLMs in software development, academic research and industry-based projects are proposing and developing various tools, benchmarks, and metrics to evaluate the efficacy of LLM-generated code. However, there is a lack of solutions and tools evaluated through empirically grounded methods, incorporating practitioners’ perspectives, to assess functionality, syntax, and accuracy of LLM-generated code in real-world applications. To address this gap, we proposed and developed a multi-model unified platform to generate and execute code based on natural language prompts. We conducted a survey with 60 software developers – practitioners from 11 countries across 4 continents working in diverse professional roles and domains - to evaluate the usability, performance, strengths, and limitations of each model. The results of this study present practitioners’ feedback and insights into the use of LLMs in real-world software development contexts, including their strengths and weaknesses, key aspects missed by benchmarks and metrics of generated code, and enhance our understanding of the practical applicability of LLMs to real-world software development. The findings of this study can inform researchers and practitioners, facilitating knowledge transfer, for a systematic selection and usage of given LLMs in software development projects. Future research focuses on integrating more diverse models into the proposed system, incorporate additional case studies, and conducting developers’ interviews for deeper empirical insights on LLM-driven software development. 

\vspace{0.2em}


The source code of proposed system, survey data, and system details can be found here: \textsf{\url{https://github.com/GPT-Laboratory/LLM-Evaluation}}

\end{abstract}
\section{Introduction}
\label{Introduction}
Large Language Models (LLMs) have shown notable performance in generating source code, acting as development bots (DevBots) to enable human-bot collaboration in software projects
 \cite{thakur2024verigen}, \cite{li2021automated}, \cite{ahmad2023towards}, \cite{rasheed2023autonomous}, \cite{rasheed2024timeless}. These models perform effectively in practical downstream tasks such as generating code from natural language descriptions \cite{nijkamp2022codegen}, \cite{sami2024early}. These advancements in LLM for complex code generation have facilitated developers with increased automation and enhanced the role of such models in software development \cite{chen2021evaluating}, \cite{roziere2023code}. For example, OpenAI's Codex, released in 2021, was designed specifically for automatic code generation \cite{chen2021evaluating}. To validate the LLM generated code, several researchers proposed benchmark tools to evaluate the accuracy and reliability of the generated code \cite{zheng2023survey}, \cite{rasheed2024large}. These tools have become key in assessing the performance of LLMs, ensuring their outputs meet the quality standards required for practical software development tasks \cite{chen2024survey}. However, there is a gap in empirical evaluation of LLM-generated code, particularly comparing their performance across diverse real-world software development tasks \cite{nguyen2023generative}.

\begin{figure}
    \centering
    \includegraphics[width=0.9\linewidth]{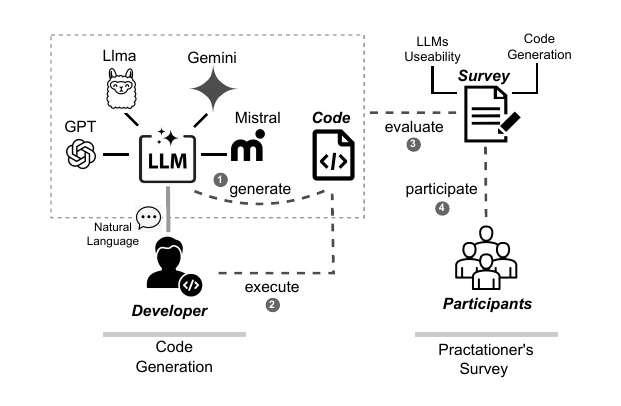}
    \caption{An overview of the proposed study}
    \label{Workflow}
\end{figure}

\textbf{Context and Motivation}:
The empirical evaluation of the functional accuracy of code generated by LLMs is an important part of our large research project, BF/Amalia-SW. This project aims to advance the automation of large-scale software development processes using LLMs, with the goal of improving code accuracy and scalability in the software development domain. In collaboration with leading Finnish companies specializing in LLM code evaluation, this project aims to conduct an empirical evaluation of eight LLMs for code generation to understand practitioners' perspectives. The motivation behind this study, to understand the strengths, limitations, and functional suitability of LLM-based code generators in real-world scenarios. Several researchers have proposed various benchmarks to evaluate the accuracy of the code generated by LLMs \cite{xu2024lvlm}. Prominent benchmarks include HumanEval \cite{chen2021evaluating}, MBPP \cite{austin2021program}, APPS \cite{hendrycks2021measuring}, and others, each designed to assess different aspects of code generation, such as functionality, complexity, and domain-specific performance. However, these benchmarks are limited by their reliance on synthetic datasets and predefined test cases, which fail to capture real-world complexity and usability concerns \cite{dou2024s}. To this end, we conducted a two-step empirical study, overviewed in Figure \ref{Workflow} to i) develop a unified system to integrate multiple LLMs to generate code based on natural language prompts, and ii) evaluate the usability, performance, strengths, and limitations of each model based on the findings of a survey with 60 software developers \footnote{The terms \textit{developer}, \textit{practitioner}, and \textit{participant} have been used interchangeably in this paper all referring to professionals engaged in this study to generate code (developer) and/or provide survey feedback (participant).}. 

\textbf{Objectives and Contributions}:
The primary objective of this study is to enable human-bot collaborative software development and to objectively assess the strengths and weaknesses of eight prominent LLMs, including Llama 3.2 3B Instruct, Mixtral 8×7B Instruct, GPT-4o, GPT-4 Turbo, GPT-3.5 Turbo, and Gemini 1.5 Flask-8B. To achieve this objective, as shown in Figure \ref{Workflow}, first, we developed a system that provides a unified platform for developers to seamlessly access above-mentioned LLMs for code generation. The developed system has been deployed as a sandbox environment to ensure a secure setting for generating and executing the code. To validate the efficacy of these LLMs, we conducted a survey with 60 developers from 11 countries across 4 continents, working in diverse professional roles and domains with experience that varied from 5 years to 20 years of coding in various programming languages. The developed system and survey questionnaire were shared with practitioners, who provided a total of 253 project descriptions, with an average of 4 descriptions contributed by each participant.  We outline the key contributions of this research as:

\begin{itemize}
    \item \textit{Implementing a multi-model system} that integrates eight LLMs to provide developers with a unified platform for comparing and evaluating code generation models on real-world systems.
    \item \textit{Conducting a survey }with 60 practitioners to incorporate practitioners perspectives, where the results offer an objective comparison of models' abilities to understand and interpret project descriptions and generate code.
    \item \textit{Providing empirically grounded guidelines} on the potential, limitations, usability, and functional efficacy of the selected LLMs as code generators.
\end{itemize}

To support replication, validation, and broader exploration of the findings of our study, we have publicly released the dataset and the proposed system via \cite{llm_evaluation} to support further research and implementation.

\vspace{0.2em}

\textbf{Structure:}  Related work is discussed in Section \ref{Background} to contextualize the scope and contributions of proposed research. The method to conduct this research is elaborated in Section \ref{Research Method}. The results of the study are presented in Section \ref{results}. Discussion of key findings, their implications, and validity threats is provided in Section \ref{Discussion}. Conclusions and vision for future research are detailed in Section \ref{conclusion}.

\section{Related Work}
\label{Background}
This section overviews the most relevant existing research on LLMs for code generation (Section \ref{LLM for code generation}) and evaluation of code generating LLMs (Section \ref{LLM evaultion}) to contextualize the scope and contribution(s) of the proposed research.

\subsection{Large Language Models for Code Generation}
\label{LLM for code generation}

Several studies have explored the potential of LLMs in supporting automated programming tasks \cite{lyu2024automatic}.
For example, AlphaCode \cite{li2022competition} reportedly improved the accuracy of code generation in programming contests, whereas Codex \cite{chen2021evaluating} enhances Copilot by offering instant coding suggestions.
Wang \textit{et al}. \cite{wang2021codet5} proposed CodeT5, a pre-trained encoder-decoder Transformer model that employs a unified approach to effectively support both code comprehension and generation tasks, while also enabling multi-task learning.
Additional open-source models for code generation include GPTNeo \cite{black2021gpt}, GPT-J \cite{wang2021gpt}), CodeParrot \cite{tunstall2022natural}, PolyCoder \cite{xu2022systematic}, CODEGEN \cite{nijkamp2022codegen}, INCODER \cite{fried2022incoder}, and CodePori \cite{rasheed2024codepori}. 
Chen \textit{et al}. \cite{chen2022codet} utilized the Codex inference API from OpenAI, alongside two open-source models, CODEGEN and INCODER, for zero-shot code generation. 
Qian \textit{et al}. \cite{qian2023communicative} proposed the ChatDev model, which integrate LLMs throughout the entire software development lifecycle. 
This model completes the full software development process in less than seven minutes at a cost of under one dollar. 
Zheng \textit{et al}. \cite{zheng2023codegeex} introduced CodeGeeX, a multilingual model with 13 billion parameters, specifically designed for code generation.
As of June 2022, CodeGeeX was pre-trained on a large dataset containing 850 billion tokens across 23 programming languages.
In recent years, Hong \textit{et al}. \cite{hong2023metagpt} proposed MetaGPT, a meta-programming framework that integrates efficient human workflows into LLM-based multi-agent collaborations. It employs an assembly line paradigm to assign diverse roles to agents, effectively breaking down complex tasks into subtasks managed collaboratively by multiple agents. Similarly, Rasheed \textit{et al}. \cite{rasheed2024codepori} introduced the CodePori system, which focuses on enhancing the understanding of complex, real-world tasks and improving the generation of accurate and efficient code for large-scale and intricate projects.

\subsection{Large Language Models Evaluation}
\label{LLM evaultion}
The evaluation of LLM-generated code has received attention from researchers, resulting in the development of benchmarks to assess various aspects of code generation across multiple programming languages \cite{chen2021evaluating}. Code generation benchmarks typically include various coding tasks
where a natural language description serves as input, and the corresponding code serves as the ground truth output \cite{ghosh2024benchmarks}. LLM-based code generation is primarily evaluated through functional correctness, which is generally determined by executing test cases to verify the expected outputs. 

HumanEval \cite{chen2021evaluating} is one of the earliest and most extensively studied benchmarks for assessing LLM-generated code. It includes 164 Python function signatures paired with docstrings and corresponding test cases to validate correctness. Austin \textit{et al}. \cite{austin2021program} introduced the MBPP benchmark, a Python-focused dataset developed through crowd-sourcing. It contains 974 programming problems, each with a description (docstring), a function signature, and three associated test cases. 

Apart from Python, other benchmarks have been developed to target additional programming languages, including Spider \cite{yu2018spider} for SQL, HUMANEVAL-X \cite{peng2024humaneval} for languages like C++, JavaScript, and Go, and MultiPL-E \cite{cassano2023multipl}, which extends HUMANEVAL and MBPP to cover 18 different programming languages. Similarly, the APPS benchmark, proposed by Hendrycks \textit{et al}. \cite{hendrycks2021measuring}, includes over 10,000 coding problems ranging from simple to complex tasks, designed to cover real-world programming scenarios and assess problem-solving skills. In addition to benchmarks like HumanEval and MBPP, other benchmarks used for code evaluation include CodeNet \cite{puri2021codenet}, which addresses diverse programming tasks across multiple languages and the CodeT5 Benchmark Suite \cite{wang2021codet5}, which evaluates tasks such as code summarization, generation, and refinement. The above mentioned benchmarks primarily evaluate function- or statement-level code. In response, Du \textit{et al}. \cite{du2023classeval} introduced ClassEval, a benchmark designed to assess class-level code generation, marking the first study to evaluate LLMs on such tasks.

To date, several code generation benchmarks have been developed using both automated and manual methods. While these benchmarks primarily emphasize lexical-based evaluation, ODEX \cite{wang2022self} introduces execution-based evaluation as a more practical approach. However, in real-world scenarios, there remains a lack of empirical evaluation of LLM-generated code, which requires practitioners to carefully test the code's functionality and adaptability in practical applications \cite{fan2023large}. Nguyen-Duc \textit{et al}. \cite{nguyen2023generative} call for more studies evaluating LLMs in industrial contexts rather than in experimental or classroom settings.

\section{Study Design}
\label{Research Method}

In this section, we explain the design of the proposed multi-model system and the survey design process. The technical details of the proposed system are provided in Section \ref{System Design}, while the survey design is discussed in Section \ref{survey design}.

\subsection{Model-based Code Generation}
\label{System Design}
\begin{figure}[t]
    \centering
    \includegraphics[width=1.0\linewidth]{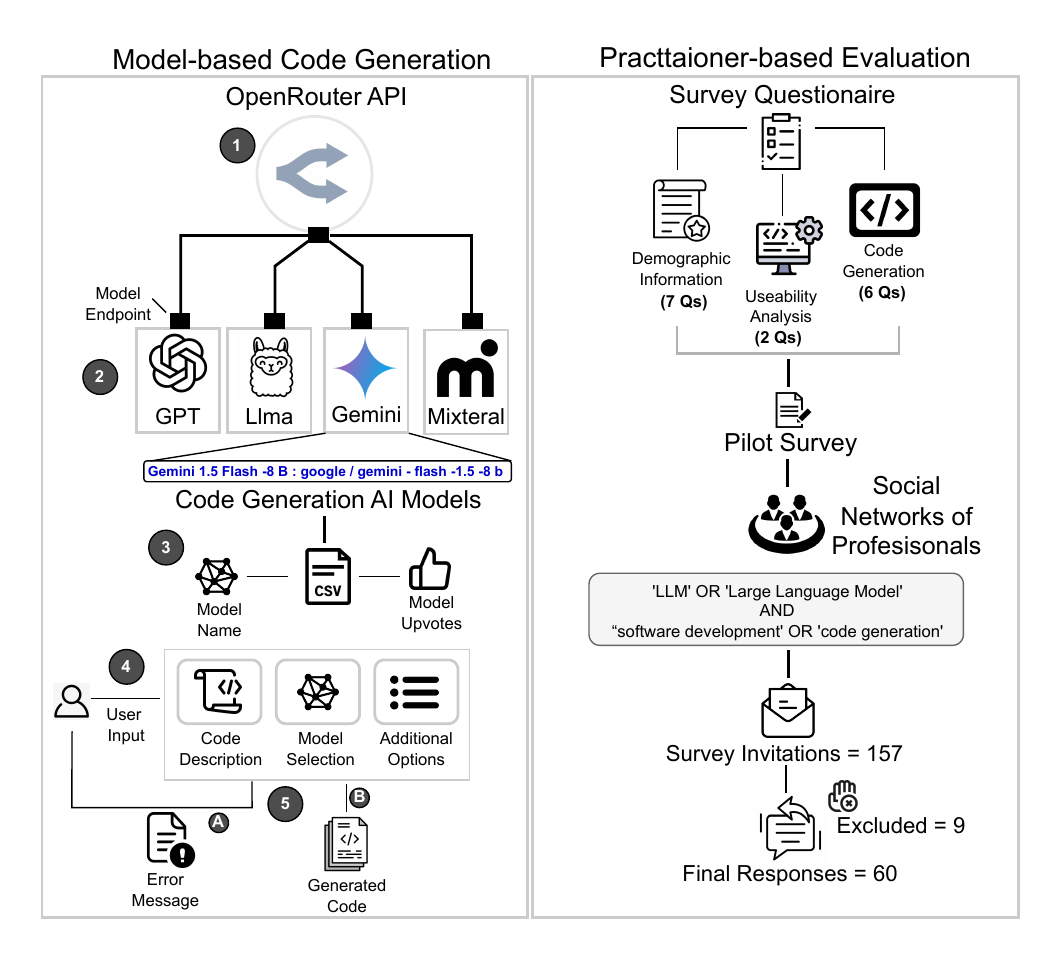}
    \caption{Workflow for solution execution and evaluation}
    \label{fig:enter-label}
\end{figure}

In this project, we integrate eight LLMs into a single platform, including Llama 3.2 3B Instruct, Mixtral 8*7B Instruct, GPT-3.5, GPT-4, GPT-4 Turbo, GPT-4o, and Google Gemini, configured exclusively for code generation. The reason for choosing the models mentioned above is their widespread adoption by practitioners for code generation \cite{roziere2023code}, \cite{elgedawy2024ocassionally}, \cite{akter2023depth}. To provide a single platform to practitioners, we utilize the OpenRouter API, which serves as the communication gateway between the application and the above-mentioned LLMs. As shown in Figure \ref{fig:enter-label}, the process is divided into three primary components: model selection, code prompt, and code generation.

\begin{itemize}
    \item \textbf{Model Selection}
Practitioners primarily select models for code generation. Each model is associated with a unique endpoint to ensure continuous compatibility with the API. By integrating multiple models from diverse providers, the system handles a wide range of computational requirements and user-specific tasks. This approach allows for dynamic selection of models based on task complexity, resource availability, and user preferences. Additionally, it ensures continuous scalability as new models can be incorporated without disrupting the existing architecture. 
\item \textbf{Code Prompts}
\label{Code Prompts}
Once the models are defined and data is prepared, the system retrieves detailed input from the user. This input includes the task description and the chosen model from the available list. These details ensure that the system has all the necessary information to process the user request effectively. To maintain operational stability, the algorithm also validates the user input, checking for completeness and correctness. If any essential parameter, such as the task description or model selection, is missing, the system promptly returns an error message to the user.
\item\textbf{Code Generation}: After the system is initialized and user inputs are validated. The system then configures the API request, specifying the selected LLM model, authorization header using the OpenRouter API key, and including the generated prompt in the request payload. This request is sent to the OpenRouter API endpoint, where it is processed to retrieve the code or an explanation. If the API request fails due to connectivity or other issues, the system logs the error and returns an error message to the user, ensuring error handling and transparency.

Finally, the system handles client interactions through HTTP requests. HTTP requests loads the model data and processes user inputs for code generation and returns the generated code. These functionalities ensure continuous interaction between the system and users, enabling efficient request handling and feedback integration. The system concludes its operation by starting the Flask server on a designated port, completing the setup for a fully functional, interactive application.
\end{itemize}

\subsection{Survey-based Model Evaluation}
\label{survey design}

Considering the research goals, we decided to conduct a survey as the primary data collection method to understand practitioners' perceptions of various LLMs for code generation. 

\subsubsection{Survey questionnaire.}
\label{creating questionnaire} 
The questionnaire was prepared in English to accommodate practitioners from various countries. It was created using the Microsoft Forms survey platform. The questionnaire included closed-ended and open-ended questions and was divided into three sections.

The first part, 7 questions about the background information of the practitioners, such as, we asked the practitioners about their demographic information such as their education, experience and their roles and level of experience in software development. Such information was expected to help us to assess the participants’ eligibility for the survey based on the inclusion and exclusion criteria defined in Table \ref{inclusion_exclusion_criteria}. In the second part, we designed two survey questions (Q8 to Q9), the aim of which is to identify the usability of the proposed system. In last part, we designed five survey questions (Q10 to Q15) in order to understand according to practitioners which models performance is best for code generation. The objective of these survey questions (Q10 to Q15) is to understand the experiences of the participants with various LLMs in the context of code generation. The main purpose is to identify the most effective model in terms of code accuracy, quality, and user satisfaction, as well as gather feedback on the usability and limitations of each model.

As observed in the questionnaire \cite{llm_evaluation}, 86.67\% of the survey questions are closed-ended questions (i.e., Q1, Q2, Q3, Q4, Q5, Q6, Q7, Q8, Q9, Q10, Q11, Q12, and Q13) with limited set of closed answer. The 13.33\% questions (i.e., Q14 and Q15) are open ended questions.  Open-ended questions are expected to enable respondents to frame their own response and avoid restriction and bias on the responses of the participants to collect qualitative data. 

\begin{table}[h!]
\centering
\caption{Inclusion and exclusion criteria for selecting valid responses}
\label{inclusion_exclusion_criteria}
\resizebox{\textwidth}{!}{%
\renewcommand{\arraystretch}{1.3} 
\begin{tabular}{|p{1cm}|p{12cm}|}
\hline
\rowcolor[HTML]{D9E2F3} 
\multicolumn{2}{|c|}{\textbf{Inclusion Criteria}} \\ \hline
\textbf{I1} & The respondent has experience in architecture design and code development. \\ \hline
\textbf{I2} & The respondent has sufficient English proficiency to answer the survey. \\ \hline
\rowcolor[HTML]{D9E2F3} 
\multicolumn{2}{|c|}{\textbf{Exclusion Criteria}} \\ \hline
\textbf{E1} & The response is not in English. \\ \hline
\textbf{E2} & The response is meaningless, e.g., it contains clear inconsistencies. \\ \hline
\end{tabular}%
}
\end{table}


\subsubsection{Evaluating and validating the questionnaire.}
\label{evaluating and validating the questionnaire}

Before uploading the survey questionnaire online for data collection, the authors conducted a review of the survey protocol to ensure consensus on the design of the survey instrument, including its objectives and questions.
Additionally, two researchers specializing in the software development field provided feedback on the survey protocol. The first author refined the survey based on the given feedback. Following these revisions, a pilot survey was conducted with five practitioners experienced in software development to further assess the reliability of the survey instrument \cite{kitchenham2008personal}. The pilot study was conducted to evaluate the clarity of the questions, the time required to complete the survey, the response rate, and the effectiveness of each question. Three participants recommended including a link to the proposed platform in the survey, along with a detailed description of its functionality and usage. Incorporating the feedback from the pilot study, a finalized version of the questionnaire was developed, comprising 15 questions that could be completed within 10–15 minutes. 

\subsubsection{Participant selection.}
\label{Participant Recruitment.}
The target population is selected based on a survey objective. We intended to investigate the software practitioners’ perspective regarding best LLM model for code generation. We did not limit the domain and countries in which the participants were working. We defined our target population as the practitioners who have experience or get involved in software development and LLMs. 

Firstly, we utilized our social networks to approach practitioners currently working in the industry with vast experience in software development and LLMs. We also used GitHub and Stack Overflow to search for the potential participants with the keyword ``large language model'' AND ``software development'' in users’ profiles. We evaluated the users on GitHub and Stack Overflow as eligible practitioners for the survey if they meet these criteria: (1) their affiliated organizations in industry were provided in their GitHub homepages or the profiles in the linked curricula vitae; and (2) their experience of software development and LLMs was described in the profiles in the linked curricula vitae. In addition, we used snowball sampling and asked participants in this survey to invite their contacts on social networks who might have been willing to participate in this industrial survey. 

After we reached the target population, we contacted the potential participants and sent them direct invitations through emails or other social media (e.g., LinkedIn, Twitter, and Facebook) depending on the types of contacts they had provided in GitHub or Stack Overflow. We sent the questionnaire to 157 potential participants. We received 27 responses with an approximate response rate of 14.65\%. Among the 23 responses, 9 invalid responses were excluded based on the criteria set for the respondents; these 9 respondents had the required experience in software development and LLMs but their answers were irrelevant. Finally, we retained 14 valid responses for data analysis. 
\subsubsection{Analyzing survey data}
\label{Analyzing valid data}
Descriptive statistics \cite{wohlin2006empirical} were used to perform a quantitative analysis of closed-ended questions. Moreover, we applied the open-coding method \cite{blair2015reflexive} to analyze open-ended response data. More details about the survey questions and the data analysis method are provided in Table \ref{data_analysis}. The first author performed open coding and then used selective coding to identify the core categories that were generated by aggregating a set of concepts. Note that we only used open coding to analyze the answers to the survey questions Q14 and Q15, since only concepts were generated from the answers. To reduce personal bias during data analysis, the other authors participated in the validation of the generated code. The disagreements were discussed and resolved. Finally, we generate 41 codes, 20 concepts, and 13 core categories from the gathered data.




\begin{table}[h!]
\centering
\caption{Relationship between survey questions and data analysis methods}
\label{data_analysis}
\resizebox{\textwidth}{!}{%
\renewcommand{\arraystretch}{1.3} 
\begin{tabular}{|l|l|l|}
\hline
\rowcolor[HTML]{D9E2F3} 
\textbf{Survey Question} & \textbf{Data Analysis Method}      & \textbf{Objective}        \\ \hline
Q1--Q7                   & Descriptive Statistic              & Demographics              \\ \hline
Q8, Q9                   & Descriptive Statistic              & Proposed System Usability \\ \hline
Q10--Q15                 & Descriptive Statistic, Open Coding & Model Performance         \\ \hline
\end{tabular}%
}
\end{table}

\section{Results}
\label{results}
In this section, we present the survey results by analyzing the responses of practitioners. First, we provide the demographic data of the practitioners, followed by an interpretation of their responses on the usability of the proposed system and the performance of LLMs in code generation.

\subsection{Demographic Data}
\label{Demographic Data}

\begin{figure}[H]
    \centering
    \includegraphics[width=0.9\textwidth]{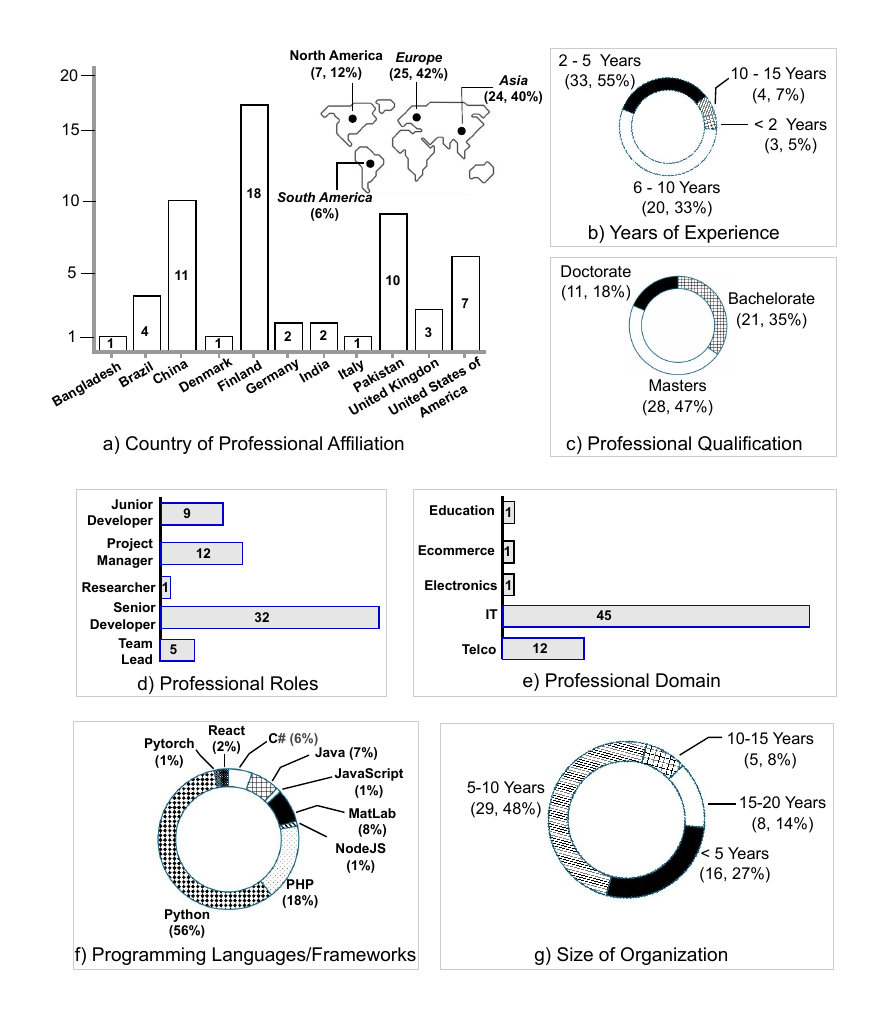}
    \caption{Demographic of data}
    \label{demographic data}
\end{figure}

To evaluate the performance of the above-mentioned LLMs for code generation, 60 practitioners participated and completed the survey. This sample size exceeds the 12 to 15 participants suggested by Guest \textit{et al}. \cite{guest2006many} as sufficient to achieve data saturation in interviews or survey research with homogeneous groups. 
The demographic details in Figure \ref{demographic data} provide insights into the participants' background, such as their experience in the specific area, the size of the company they are affiliated with, their country, and their roles, among other factors. Below, we provide the details of the demographic data to offer a clearer understanding of the participants' characteristics.

\begin{itemize}
    \item \textbf{Geo-distribution} indicates the diversity of participants based on their geographical locations. As shown in Figure \ref{demographic data} (a), the 60 respondents reside in eleven different countries, with the majority located in countries where English is an official language. Of the participants, 25\% were from Europe, 24\% from various countries in Asia, and 7\% from North America.
\end{itemize}

\begin{itemize}
    \item \textbf{Experience} of participants is important for a survey, as it ensures that responses are informed and reliable, contributing to the credibility of the findings. As shown in Figure \ref{demographic data} (b), the majority of respondents (33.55\%) have approximately 5 years of experience in software development. Additionally, 20.33\% have between 6 to 10 years of experience, and 4\% have more than 10 years of experience. Most participants (45\%) are affiliated with the IT industry, while 12\% are associated with the telecommunications industry. Furthermore, more than half of the participants hold roles as senior developers. These findings provide confidence that our respondents have the necessary experience to complete the survey questionnaire.
\end{itemize}

\begin{itemize}
    \item \textbf{Qualification} of participants is important to get well-informed responses. As illustrated in Figure \ref{demographic data} (c), the majority of respondents (28.47\%) hold a Master’s degree, 11.18\% hold a Doctorate degree, and 21.35\% hold a Bachelor’s degree.
\end{itemize}

\begin{itemize}
    \item \textbf{Professional roles} indicate the participants’ expertise in software development and LLMs. The responses of the practitioners identified a total of five roles highlighted in Figure \ref{demographic data} (d). The majority of participants (33\%) identified themselves as Senior Developers, reflecting a strong representation of experienced professionals contributing to the field. Following this, 12\% of the participants are Project Managers, while 9\% are Junior Developers. Finally, only 1\% of the participants identified as Researchers, highlighting limited academic research engagement in this survey.
\end{itemize}



\subsection{Usability and User Experience in Model-Based Code Generation}
\label{Usability of the Proposed System}

To evaluate the usability of the proposed system, a survey was conducted involving 60 practitioners who were given access to a platform that allowed them to select a model for generating code based on a given description. The survey focused on two key usability aspects: ease of use and satisfaction of expectations.

\begin{itemize}
    \item \textbf{Ease of Use}: Out of the 60 participants, 53 practitioners (92\%) stated that the proposed system was easy to use. They appreciated the intuitive interface and simple workflow, which allowed them to quickly select a model and generate code without requiring extensive technical knowledge. Some participants specifically mentioned that step-by-step instructions and user-friendly navigation significantly reduced the learning curve. On the other hand, 8 practitioners (8\%) found the system challenging to use. The primary reasons cited included occasional difficulty in understanding error messages and the lack of explanations for certain advanced features. These participants suggested that adding more detailed guidance or tutorials could make the system more accessible for first-time users.
    \item \textbf{Meeting Expectations for Code Generation} Most of the participants (approximately 92\%) indicated that the system met their expectations for code generation. They praised its accuracy in interpreting descriptions and producing functional code. Some highlighted that the ability to choose between models improved their experience as it allowed them to explore and compare outputs from different approaches. However, a small percentage (about 8\%) felt that the system occasionally did not meet their expectations. The most common issue mentioned was that the generated code sometimes required additional fine-tuning to work continuously in their specific projects. These users recommended improving the system’s contextual understanding to produce more tailored results.
    \end{itemize}

\begin{takeawaybox}{Takeaways}
\scriptsize
\graycircle{3} \textbf{Usability and Experience}: An overwhelming majority, i.e., 92\% of developers found the system easy to use in terms of model selection, code generation, and enhanced learning. On the other hand, 8\% of developers faced challenges related to unclear error messages and the lack of features for contextual understanding of code generation.\\
\graycircle{3} The majority highlighted a positive experience in selecting the most suitable model(s) based on given natural language prompts and enabling them to compare outputs from different LLMs.
\end{takeawaybox}

\subsection{LLM Performance Analysis for Code Generation}
\label{LLM Performance Analysis for Code Generation}




\subsubsection{Programming Languages and Project Description.}
Two survey questions (Q10 and Q11) were used to analyze programming language preferences and the number of project descriptions provided for code generation. As shown in Figure \ref{demographic data} (f), Python is the most widely used language, representing 56\% of the responses, highlighting its dominance due to its simplicity and extensive AI and machine learning libraries. PHP (18\%) and MATLAB (8\%) follow, indicating their relevance in specific use cases. Languages like Java (7\%) and C\# (6\%) demonstrate their continued utility in enterprise-level applications.

Practitioners' feedback highlighted several reasons why GPT-4o emerged as the preferred model, 
particularly in Python development.   
Four practitioners noted that GPT-4o's ability to understand and interpret project descriptions is superior to other models. For example, one practitioner mentioned, \textit{``GPT-4o generates relevant code with minimal manual adjustments''}. 11 practitioners believed that code generated by GPT-4o followed best practices in term of syntax and structure. For example for Python development, GPT-4o excelled in handling edge cases and producing clean, maintainable code. As a developer mentioned, \textit{``The outputs of GPT-4o were more reliable and often required less debugging compared to other models, specifically for Python language''}.

\begin{takeawaybox}{Takeaways}
\scriptsize
\graycircle{3} \textbf{Core Finding}: Python was the most preferred programming language (56\%) due to its simplicity and AI libraries, followed by PHP (18\%) and MATLAB (8\%). Practitioners favored GPT-4o for its superior understanding of project descriptions, compliance with best practices, and reliability, especially in Python development, with minimal debugging required.\\
\end{takeawaybox}

\subsubsection{Accuracy and Quality of Code Generation.}
\begin{figure}[H]
    \centering
    \includegraphics[width=1.0\textwidth]{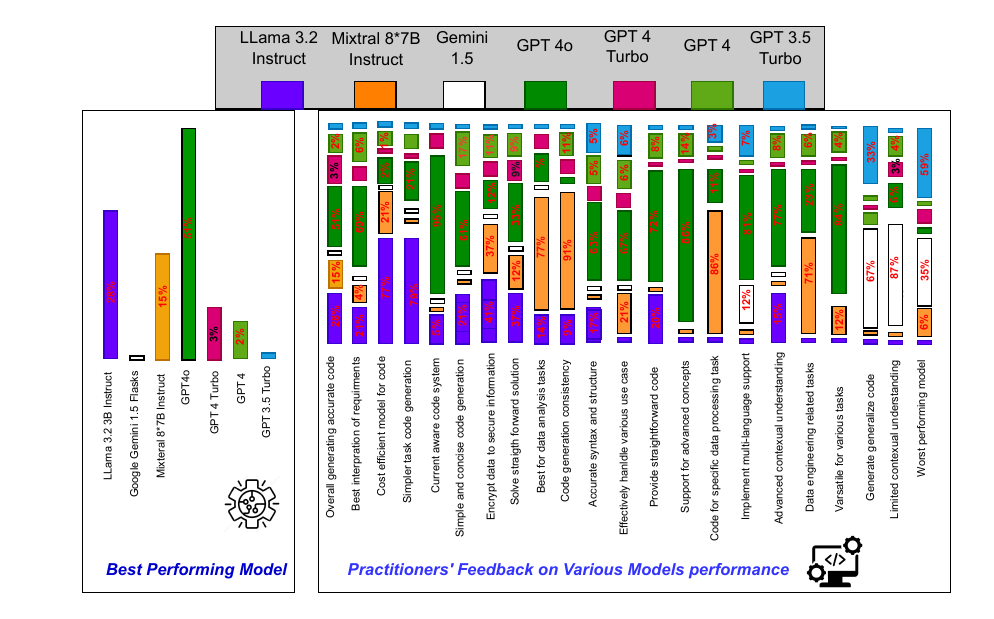}
    \caption{Practitioners' feedback on the performance of models for code generation}
    \label{practitioners feedback}
\end{figure}

We used three survey questions (Q12, Q13, Q14) to identify the best LLM for code generation in terms of accuracy and quality. As illustrated in Figure \ref{practitioners feedback}, GPT-4o emerged as the top choice, with 31 out of 60 practitioners rating it as the best model for code generation. (17 out of 60) preferred Llama 3.2 3B Instruct, and 9 considered Mixtral 8*7B Instruct the best-performing model for code generation. 

(7 out of 31) practitioners emphasized that GPT-4o effectively handles various use cases, such as generating back-end APIs, front-end components, and scripts for automation, consistently delivered accurate results. For instance, a practitioner mentioned that \textit{``I tested all the models for different tasks, but GPT-4o handled both front-end and back-end requirements perfectly, which saved a lot of time''}. (5 out of 31) practitioners mentioned that GPT-4o easily generates code that integrates the appropriate libraries and utilizes them efficiently to solve problems. As one respondent noted, \textit{``When I used GPT-4o for generating Python scripts, it imported the right libraries, which used them efficiently to solve problems. Llama 3.2 3B Instruct, Mixtral 8*7B Instruct, and others models sometimes missed these details''}. (4 out of 31) respondents highlighted that the interface and workflow provided by the system using GPT-4o were straightforward and user-friendly. For instance, one representative mentioned, \textit{``switching between models was easy, but GPT-4o stood out for how well it integrated with the overall workflow. I could generate and test code quickly without much hassle''}.

While GPT-4o emerged as the preferred choice among most practitioners, Llama 3.2 3B Instruct was identified as the second most preferred model for code generation. (17 out of 60) argue that Llama 3.2 3B Instruct often generate simpler, more concise code. (4 out of 17) practitioners highlighted that for smaller project or tasks Llama generate accurate code becuase it did not require highly complex logic. As one practitioner stated, \textit{``I liked Llama 3.2 because the code it generated was clean and easy to follow, especially for straightforward problems''}. Three respondents among the 17 argued that Llama 3.2 3B Instruct is a cost-effective and budget-friendly alternative to other models, especially for teams with limited resources. As one practitioner remarked, \textit{``LLaMA struck a balance between cost and performance, making it ideal for smaller projects where budget is a concern''}. (4 out of 17) practitioners found Llama 3.2 3B Instruct is more efficient in terms of speed. It generated outputs faster compare to other language model. For instance, a practitioner mentioned that \textit{``As I tested all the models, Llama was faster compared to the others. Its response time was within a second, whereas other models required a longer wait to generate code.''}.

(7 out of 60) practitioners selected Mixtral 8*7B Instruct as the best model for code generation. Those who preferred it highlighted several unique advantages offered by the model. For example, (4 out of 7) practitioners noted that Mixtral 8*7B Instruct performed well in certain scenarios, such as generating code for data analysis or machine learning workflows. One practitioner stated, \textit{``Mixtral worked best when I needed code for specific data processing tasks. Its outputs were accurate for these domains''}. Mixtral 8*7B Instruct was also praised for its consistency in generating outputs. (2 out of 7) participants mentioned that Mixtral 8*7B Instruct’s code generation was stable and predictable. For instance, a practitioners noted, \textit{``Mixtral might not be as advanced as GPT-4o, but its outputs were consistent, which is something I valued for repetitive tasks''}.


\begin{takeawaybox}{Takeaways}
\scriptsize
 \graycircle{3} \textbf{Core Finding}: GPT-4o emerged as the top choice for code generation, preferred by 52\% of practitioners for its accuracy, use of best practices, and ability to handle diverse tasks. It was praised for generating reliable Python code, efficiently integrating libraries, and offering a user-friendly workflow compared to other models.\\
 \graycircle{3} \textbf{Core Findings}: Llama 3.2 3B Instruct was the second most preferred model for code generation, chosen by 28\% of practitioners. It was praised for generating simpler, concise code ideal for smaller tasks, its cost-effectiveness for budget-constrained teams, and faster response times compared to other models.\\
 \graycircle{3} \textbf{Core Findings}: Mixtral 8*7B Instruct was selected by 12\% of practitioners as the best model for code generation, particularly excelling in data analysis and machine learning workflows. It was praised for its consistency and stability in generating accurate outputs, especially for domain-specific and repetitive tasks.
\end{takeawaybox}

\subsubsection{Worst Model for Code Generation.}
Survey question 15 was used to identify practitioners' perspectives on the least effective model for code generation. GPT-3.5 Turbo was rated the least favorable model for code generation by the majority of practitioners. Practitioners found that it required significantly more effort to correct and optimize its outputs, making it impractical for real-world use. Several limitations were consistently highlighted in practitioners feedback. 

Practitioners found that GPT-3.5 Turbo struggled to handle modern coding practices and frameworks. While it could generate basic code snippets, but often failed to utilize the latest libraries and conventions effectively. One practitioner noted, \textit{``the code generated by GPT-3.5 Turbo often depend on outdated methods, which need to rewrite to bring it up to standard''}.
The outputs generated by GPT-3.5 Turbo frequently contained errors, particularly when handling complex tasks. Practitioners noted that the model often misinterpreted input requirements and produced code that failed to compile correctly, highlighting its limitations in addressing intricate coding challenges. One participant shared, \textit{``I had to spend more time debugging GPT-3.5 Turbo’s outputs than writing the code myself. It was frustrating and inefficient''}.
Practitioners pointed out that GPT-3.5 Turbo struggled with maintaining context in longer or more complex project descriptions. This led to incomplete or incorrect code generation. 
A common complaint was that GPT-3.5 Turbo often generated unnecessarily lengthy and inefficient code. Practitioners stated that this made the code harder to maintain and required additional effort to optimize. For instance, one participant explained, \textit{``The code generated by GPT-3.5 Turbo was bloated and didn’t follow best practices. It required a lot of cleaning up to make it usable''}.
Practitioners noted that GPT-3.5 Turbo frequently failed to generate correct or optimal code involving advanced Python libraries. One respondent said, \textit{``Whenever I used GPT-3.5 Turbo for machine learning tasks, it either missed key steps or produced code that wasn’t functional. It was just unreliable''}.

\begin{takeawaybox}{Takeaways}
\scriptsize
\graycircle{3} \textbf{Core Finding}: GPT-3.5 Turbo was rated the least effective model for code generation due to its outdated methods, frequent errors, and inability to handle complex tasks or modern coding practices. Practitioners found its outputs inefficient, often requiring significant debugging and optimization, making it impractical for real-world use.\\
\end{takeawaybox}

\section{Discussion} 
\label{Discussion}
In this section, we summarize and interpret the main findings of our study. We discuss their implications for practitioners and researchers.

\subsection{Summary of the Main Findings}

This study evaluates the performance of eight prominent LLMs for code generation through an empirical investigation that involved 60 practitioners. Our findings show that GPT-4o emerged as the most effective model, with 31 practitioners favoring its ability to generate accurate, functional, and context-aware code, particularly for Python-based tasks. Llama 3.2 3B Instruct was identified as the second-best model, particularly suitable for smaller and less complex projects. Mixtral 8*7B Instruct, although less preferred overall, was noted for its effectiveness in specific domains such as data analysis and machine learning. In contrast, GPT-3.5 Turbo was ranked as the least effective model, as it required substantial post-generation corrections and often produced outdated code.

\subsection{Implications for Researchers and Practitioners}
\subsubsection{For researchers:} they integrate the identified strengths and weaknesses of these models to develop targeted improvements, such as enhancing contextual understanding and usability for complex tasks. The results of this study provide researchers with valuable evidence to guide the future fine-tuning of the above-mentioned LLMs for code generation, ensuring improved performance and applicability in real-world scenarios. Furthermore, the evaluation methodology outlined in this paper can serve as a replicable framework to evaluate future LLMs \cite{llm_evaluation}.

\subsubsection{For practitioners:} this study offers a practical understanding of the capabilities and limitations of various LLMs in real-world software development scenarios. The results suggest that GPT-4o can be an ideal choice for generating high-quality and maintainable code, while Llama 3.2 3B Instruct offers a cost-effective alternative for smaller tasks. Practitioners can also benefit from integrating model feedback mechanisms, as demonstrated in this study, to refine the selection of models for specific use cases.

\subsection{Threat to Validity of the Study} 
\label{Threat to Validity}
We follow the guidelines provided by Runeson et al. \cite{runeson2009guidelines} to address the threats to the validity of our study. Several factors represent potential threats to the validity of the study results. However, we did not address internal validity, as this study does not explore causal relationships.
\subsubsection{Construct validity}
\label{Construct Validity}
is the extent to which a measurement accurately represents the theoretical concept it is intended to measure \cite{runeson2009guidelines}.
In this study, construct validity depends on the accuracy and clarity of the survey questionnaire, as well-designed questions can logically support the results, whereas poorly designed questions may affect the survey outcomes. To address this threat, we developed clear and well-defined questions for the questionnaire, based on the proposed system. All the questions were reviewed by the second and third authors, and any disagreements were discussed between them. Specifically, open-ended questions in the survey could be misinterpreted or lead to irrelevant or unclear responses. To address this, we conducted a pilot study by sharing the survey with five practitioners, whose feedback helped us refine and enhance the questionnaire.
Another threat arises from whether participants provide truthful answers, as their responses may differ from their actual thoughts. To minimize this threat, we utilized history logs in our proposed system to record how many project descriptions were provided by practitioners to validate the performance of the above-mentioned LLMs. In the invitation letter, we confirmed that participant names and project description data would remain anonymous, ensuring that no respondent identities would be disclosed.

\subsubsection{External validity}
\label{External Validity}
is the degree to which the findings can be generalized to other contexts \cite{runeson2009guidelines}. In this study, we evaluate the performance of eight prominent LLMs for code generation from the perspective of practitioners with experience in software development and working with LLMs. Demographic data indicate that most of the participants work on medium-sized projects, which can introduce the risk of an unbalanced sample. Therefore, a primary threat to this study is whether the 60 respondents accurately represent software practitioners. We carefully analyzed the participants' profiles on GitHub and Stack Overflow and were satisfied that the respondents represented a diverse background and were suitably qualified to participate in this study. This makes us confident that the results are presentative and generalizable to some extent. Another threat is the relatively small response rate, as we received only 60 valid responses, i.e, 
response rate was 10.3\%.
There is no explicit standard or guideline for 
an acceptable response rate for software engineering survey studies, apart from self-reported data provided by some researchers. For instance, Chakraborty \textit{et al}. \cite{chakraborty2018understanding} reported a response rate of approximately 13\% in their survey.

\subsubsection{Reliability}
\label{Reliability}
refers to the replicability of a study in producing the same results \cite{runeson2009guidelines}. There is a potential threat that other researchers replicate our study and generate different results. We mitigated this threat by offering a comprehensive description of our research method.
Furthermore, to ensure the reliability of our results and findings, we have made all data from this study publicly available in \cite{llm_evaluation}, enabling other researchers and practitioners to validate our results. The measures outlined above can help improve the reliability of the results in our study.

\section{Conclusion}
\label{conclusion}
In this study, we conducted an empirical evaluation of eight prominent LLMs for code generation, addressing the need for empirical evaluations grounded in practical use cases. We initially proposed a unified system to integrate the LLMs mentioned above into a single platform. Then we conducted a survey with 60 practitioners by providing the unified system along with the survey questionnaire. The feedback of the practitioners highlighted the strengths and limitations of each model, providing important insights into aspects that were not captured by existing benchmarks and metrics. The result indicates that the ability of GPT-4o to understand and interpret project descriptions is better than other models. By sharing our findings, we aim to support the research community in further refining LLMs for code generation and related applications.

\textbf{Future work} is focused on integrating more models into a unified platform and increasing the number of practitioners to capture broader industrial perspectives. This includes evaluating the models' performance in domain-specific contexts and refining the system to better address real-world software development challenges. 
These efforts aim to improve the applicability of LLMs by aligning their capabilities more closely with industry needs.
%


%
%
\bibliographystyle{splncs04}
\bibliography{sn-bibliography}
%




\end{document}